\documentclass[12pt]{article}
\usepackage{amssymb}
\usepackage{bm}
\usepackage[utf8]{inputenc}
\usepackage[english]{babel}
\usepackage{xcolor}
\begin{document}
\title{Lifetime predictions for virgin and recycled 
high-density polyethylene under creep conditions}

\author{A.D. Drozdov,\footnote{E--mail: aleksey@m-tech.aau.dk}
\hspace*{1 mm} R. H{\o}j Jermiin, and J. deClaville Christiansen\\
Department of Materials and Production\\
Aalborg University\\
Fibigerstraede 16, Aalborg 9220, Denmark}
\date{}
\maketitle

\begin{abstract}
Recycling has become a predominant subject in industry and science 
due to a rising concern for the environment driven by high production 
volume of plastics.
Replacement of virgin polymers with their recycled analogs
is not always possible because recyclates cannot met the same 
property profiles as their virgin counterparts.
To avoid deterioration of the mechanical properties, it is proposed 
to replace a virgin polymer with a recycled polymer of another grade 
whose characteristics (measured in tensile tests) are close to those of 
the virgin material.
This approach opens a way for the use of recycled polymers 
in short-term application, but its suitability for long-term applications 
has not yet been assessed.
A thorough experimental investigation is conducted of the mechanical 
response of virgin high-density polyethylene (HDPE) used for insulation 
of pipes and recycled HDPE manufactured from post-consumer plastic waste
(their stiffness, strength and elongation to break adopt similar values).
A model is presented in viscoelastoplasticity of semicrystalline
polymers. 
Its parameters are determined by matching experimental data
in short-term relaxation and creep tests.
The lifetime of virgin and recycled HDPE under creep conditions is
evaluated by means of numerical simulation.
It is shown that the stress--time to failure diagrams for virgin and 
recycled HDPE practically coincide.
\end{abstract}

\noindent
{\bf Key-words:} High-density polyethylene; Secondary recycling; Creep rupture; 
Lifetime assessment; Modeling.

\section{Introduction}

Although polymers have become an indispensable part of our life,
an extreme increase in their production has lead to a steadily growing
proportion of plastic waste going into landfill. 
Due to considerable environmental concerns, recycling and reuse of 
polymers has become a predominant subject in industry and science
\cite{ITB14,WCH22,LN22}. 

Four approaches to recycling of plastic wastes are conventionally
distinguished \cite{TS20}:
\begin{enumerate}
\item
At primary recycling, clean, uncontaminated, single-type waste
with a controlled history is blended in-plant with virgin polymers.

\item
At secondary recycling, plastic waste is sorted, separated from contaminants
and other waste materials, cleaned, and compounded, before being mixed
with virgin polymers.

\item
Under tertiary recycling, solid plastic materials are converted into 
smaller molecules by means of chemical treatment.
These intermediates are used as feed stocks for the production 
of new plastics.

\item
The quaternary recycling consists in clean incineration 
(thermal and/or catalytic pyrolysis) to recover the energy content 
of plastic wastes.
\end{enumerate}

This study focuses on the secondary recycling of high-density polyethylene
(HDPE) and application of recycled HDPE for production of pipes.
Due to degradation of post-consumer plastics during 
their service life and their thermo-mechanical degradation 
induced by recycling processes \cite{KPP99,KPP00},
mechanical characteristics of recycled polymers, 
in particular, polyolefins, 
are lower than those of their virgin analogs \cite{VK08}.
Although deterioration of the mechanical properties 
observed in short-term tests on recycled HDPE 
is not very pronounced compared to the virgin polymer
\cite{OGZ15,CSV21,VPM21}, 
it leads to a catastrophic reduction in the lifetime measured
in long-term experiments.
For example, the lifetime under creep conditions is reduced by
one to two orders of magnitude \cite{APB15,SSM22},
the number of cycles to failure under fatigue conditions 
decreases by an order of magnitude \cite{ZHR23a,ZHR23b},
and time to failure caused by crack propagation diminishes by
one to three orders of magnitude \cite{SY00,NNS18,JDR20}.

To ensure comparable long-term properties of virgin and recycled plastics,
it appears natural to replace a virgin HDPE with a recycled polymer
(prepared from a different waste feedstock) whose mechanical properties after 
recycling are close to those of the virgin material.
This approach has recently been proposed in \cite{IJM21,JDR21} 
where its environmental and economical advantages are discussed.
The objective of this study is to demonstrate that 
HDPE samples manufactured from virgin and recycled polymers
with similar mechanical characteristics (the Young modulus, yield stress,
and strain at break) have practically identical lifetimes under
creep conditions (which guarantees that they withstand 
50 years of service life under tensile creep with the same stresses).
This result is rather unexpected because creep resistance of solid 
polymers is determined to a large extent by their viscoelastic and 
viscoplastic properties (which differ noticeably for the virgin and
recycled HDPE).

For the analysis, two grades of high-density polyethylenes were chosen:
a virgin HDPE used for insulation of pipes for district heating
and a recycled HDPE manufactured from post-consumer household plastic waste.
The similarity of their mechanical responses was confirmed by observations
in uniaxial tensile tests with a constant cross-head speed.
A thorough investigation of the viscoplastic and viscoplastic properties
of these polymers was conducted in short-term tensile relaxation tests with 
various strains and creep tests with various stresses.
Predictions of their lifetime in long-term uniaxial creep tests was performed 
with the help of the model recently developed in \cite{DHC23} and validated
by comparison of observations in short-term tests with results of 
simulation.

The exposition is organized as follows.
Experimental data on recycled HDPE in mechanical tests are reported 
in Section 2, where they are compared with observations on
virgin HDPE presented in \cite{DHC23}.
A brief description of the model in viscoelastoplasticity of
semicrystalline polymers is provided in Section 3.
Fitting of experimental data is conducted in Section 4.1.
Validation of the model is performed in Section 4.2
where results of simulation are compared with observations.
In Section 4.3, the stress--time to failure diagrams are discussed for 
virgin and recycled HDPE under tensile creep conditions.
Concluding remarks are formulated in Section 5.

\section{Materials and Methods}

As a virgin material, we chose HDPE Borsafe HE3490-LS 
purchased from Borealis.
It has density 959 kg/m$^{3}$, 
melt flow rate 0.25 g/10 min, 
and melting temperature 130 $^{\circ}$C.
For comparison, we used high-density polyethylene PEHD-R-E-GREY 
supplied by Aage Vestergaard Larsen A/S (Denmark). 
This polymer is manufactured from post-consumer household plastic waste,
and it has density 960 kg/m$^{3}$, 
melt flow rate 1.2 g/10 min,
and melting temperature 130 $^{\circ}$C. 

Dumbbell-shaped specimens (ISO 527-2-1B) with the total length 145 mm, 
the gauge length 65 mm, and the cross sectional area 9.81 mm $\times$ 3.95 mm
were prepared by using injection-moulding machine Ferromatik Milacron K110. 

Uniaxial tensile tests were performed by means of the testing machine
Instron 5568 equipped with an extensometer and a 5 kN load cell.
Each test was repeated at least by twice.

Three series of tests were performed: 
(i) tensile tests with cross-head speeds $\dot{d}=50$ mm/min 
up to breakage of samples, 
(ii) short-term stress relaxation tests (with the duration of 30 min) 
with strains $\epsilon$ ranging from 0.01 and 0.05,
and 
(iii) short-term creep tests (with the durations up to 6 h)
with various stresses $\sigma$ belonging to the interval between 9 and 20.5 MPa.
All experiments were conducted at room temperature ($T=21$ $^{\circ}$C)
in a climate-controlled room.

\subsection{Tensile tests}

Uniaxial tensile tests on virgin and recycled HDPE were conducted 
with cross-head speed $\dot{d}=50$ 
(which corresponded to the strain rate $\dot{\epsilon}=0.0088$ s$^{-1}$)
until breakage of samples. 
This cross-head speed is conventionally used for quality assessment 
of HDPE pipes under quasi-static loading \cite{NCM22}.
The tests were repeated five times on samples of virgin HDPE and 
six times on samples of recycled HDPE.
The specimens were prepared by injection molding under 
the same conditions.

\input{figure-01.tex}

Experimental data on virgin and recycled HDPE are depicted in Figure 1, 
where the engineering tensile stress $\sigma$ is plotted versus 
the engineering strain $\epsilon$.
For both polymers under consideration, 
the stress $\sigma$ increases with strain $\epsilon$ 
below the yield point $\epsilon_{\rm y}$, 
reaches its maximum (the yield stress $\sigma_{\rm y}$) at $\epsilon_{\rm y}$,
and decreases slightly afterwards.
The strain at break $\epsilon_{\rm b}$ is calculated as the maximum 
strain along the stress--strain diagram.
Figure 1 reveals good repeatability and reliability of measurements.
For example, the standard deviation $\delta_{E}$ of the Young's modulus $E$
and the standard deviation $\delta_{\sigma_{\rm y}}$ of the yield stress 
$\sigma_{\rm y}$ do not exceed 1\% of their average values $\bar{E}$ and
$\bar{\sigma}_{\rm y}$, respectively.
The standard deviation $\delta_{\epsilon_{\rm b}}$ of
the strain at break $\epsilon_{\rm b}$ is relatively large,
which may be explained by significant toughness of samples.
The average values and the standard deviations of parameters $E$,
$\sigma_{\rm y}$ and $\epsilon_{\rm b}$ are collected in Table 1.
This table shows that the stiffness (characterized by $\bar{E}$) 
of the recycled polymer exceeds that of the virgin 
HDPE by 35\%, their strengths (estimated by $\bar{\sigma}_{\rm y}$) 
coincide practically, whereas the toughness of the recycled HDPE
(characterized by $\bar{\epsilon}_{\rm b}$) is lower than
that of virgin HDPE by 25\%.

\noindent
{\bf Table 1:} Material parameters for virgin and recycled HDPE determined
in tensile tests.
\vspace*{3 mm}

\begin{center}
\begin{tabular}{@{} l c c c c c c @{}}\hline
 & $\bar{E}$ GPa &  $\delta_{E}$ GPa & $\bar{\sigma}_{\rm y}$ MPa  & $\delta_{\sigma_{\rm y}}$  MPa & $\bar{\epsilon}_{\rm b}$ & $\delta_{\epsilon_{\rm b}}$ 
\cr \hline
Virgin   & 0.966      & 0.072             & 25.429      & 0.216     &  0.397 & 0.180  \\
Recycled & 1.305      & 0.014             & 26.166      & 0.224     &  0.297 & 0.073  \\
\hline
\end{tabular}
\end{center}

\subsection{Relaxation tests}

Short-term stress relaxation tests on recycled HDPE were performed 
at strains $\epsilon_{0}=0.01$, 0.02 and 0.05.
Tests were repeated by twice on different specimens manufactured
under the same conditions.
In each test, a specimen was stretched  with a cross-head speed 
of 50 mm/min up to the required strain $\epsilon_{0}$.
Then, the strain was fixed, and the decay in stress $\sigma$
was measured as a function of relaxation time $t_{\rm rel}=t-t_{0}$,
where $t_{0}$ stands for the instant when the strain $\epsilon$ reaches 
its ultimate value $\epsilon_{0}$.

\input{figure-02.tex}

Experimental data in relaxation tests on recycled HDPE 
are reported in Fig. 2 together with their fits by the model.
In this figure, the engineering stress $\sigma$ is plotted versus 
the logarithm ($\log=\log_{10}$) of relaxation time $t_{\rm rel}$.
Circles stand for the experimental data, and 
solid lines denote their fits by the model described in Section 3.

\subsection{Creep tests}

Two series of uniaxial creep tests on recycled HDPE were performed 
with ``low" ($\sigma=9.0$, 10.0, 12.0, and 14.5 MPa)
and ``high" ($\sigma=17.5$, 18.0, 19.0, 20.0 and 20.5 MPa)
tensile stresses.
Each test was repeated by twice on specimens prepared under
the same condition.
In a creep test, a sample was stretched up to the required stress $\sigma$
with a cross-head speed of 50 mm/min.
Then, the stress was fixed, and an increase in tensile strain $\epsilon$
was measured as a function of creep time $t_{\rm cr}=t-t_{0}$,
where $t_{0}$ stands for the instant when the required stress $\sigma$ 
is reached.
The duration of creep tests with low stresses (6 h) was fixed.
Creep tests with high stresses were conducted until breakage of samples.

\input{figure-03.tex}

\input{figure-04.tex}

Experimental data in creep tests on recycled HDPE
are reported in Figures 3 (low stresses) and 4 (high stresses).
In these figures, tensile strain $\epsilon$ is plotted versus creep 
time $t_{\rm cr}$.
For each stress $\sigma$, circles denote the mean (over observations 
on two samples) values of tensile strain,
and the solid line stands for their approximation by the model described
in Section 3.

To assess the repeatability of measurements and the accuracy  
of their fitting, experimental data in creep tests with 
two highest stresses ($\sigma=20.0$ and 20.5 MPa) are reported
in Figure 5 together with results of numerical simulation.
This figure reveals good reproducibility of observations.
Figures 4 and 5 show that for all stresses under consideration,
transition to the stage of tertiary creep (characterized by 
a rapid growth of tensile strain induced by damage of samples)
occurs at the critical strain $\epsilon_{\ast}$ close to 0.4. 

\input{figure-05.tex}

Experimental data in relaxation and creep tests on virgin HDPE
are reported in \cite{DHC23}. 
Comparison of observations on virgin and recycled HDPE shows
similarities of their time-dependent behavior.
A detailed discussion of differences between their responses 
is provided in Section 4.

\section{Model}

The viscoelastoplastic response of HDPE under isothermal deformation 
with small strains is described by a constitutive model based on
the following assumptions.

HDPE is a semicrystalline polymer consisting of two 
(amorphous and crystalline) phases 
(for simplicity, the presence of the rigid amorphous phase 
and inter-phases is disregarded \cite{BDC08}).
The viscoplastic deformation of the amorphous phase describes 
(i) chain slip through the crystals,
(ii) sliding of tie chains along and their detachment from lamellar
blocks, and (iii) detachment of chain folds and loops from surfaces
of crystal blocks \cite{HHL99}.
The viscoplastic deformation in the crystalline phase reflects 
(i) inter-lamellar separation, (ii) rotation and twist of lamellae,
and (iii) fine and  coarse slip of lamellar blocks \cite{GBA92}.
The viscoelastic response is associated with 
(i) relaxation of stresses in chains located in the amorphous regions
and (ii) time-dependent decay in forces transmitted 
to the crystalline skeleton by tie chains \cite{DG03a}.

A detailed account for the evolution of micro-structure in
semicrystalline polymers under loading leads to a strong increase
in the number of adjustable parameters.
To make the model tractable, we treat HDPE as an equivalent 
non-affine network of polymer chains connected by permanent 
and temporary junctions \cite{DC08}. 
The non-affinity of the network means that junctions between chains
can slide with respect to their reference positions under deformation.
Keeping in mind that the sliding process reflects viscoplastic 
deformations in the amorphous and crystalline regions, 
the strain tensor for plastic deformation 
$\bm{\epsilon}_{\rm p}$ consists of two components
\[
\bm{\epsilon}_{\rm p}=\bm{\epsilon}_{\rm pa}+\bm{\epsilon}_{\rm pc},
\]
where the strain tensors $\bm{\epsilon}_{\rm pa}$ and 
$\bm{\epsilon}_{\rm pc}$ describe viscoplastic deformations in
the amorphous phase and the crystalline skeleton, respectively. 
A similar decomposition of the plastic strain into two
components was introduced in \cite{DKC13} to describe
the viscoplastic deformation of semicrystalline polymers at finite strains.

Chains in the polymer network are bridged by permanent and temporary bonds.
Both ends of a permanent chain are merged with the network
by permanent bonds.
At least one end of a temporary chain is connected with the network
by a temporary (physical) bond that can break and reform.
When both ends of a temporary chain are attached to the network,
the chain is in its active state.
When an end of an active chain separates from its junction at some
instant $\tau_{1}$, the chain is transformed into the dangling state.
When the free end of a dangling chain merges with the network
at an instant $\tau_{2}>\tau_{1}$, the chain returns into the active state.
Attachment and detachment events occur at random times being
driven by thermal fluctuations \cite{TE92}.

The network consist of meso-regions with various 
activation energies for breakage of temporary bonds.
The rate of separation of an active chain from its junction
in a meso-domain with activation energy $u$ is governed by the Eyring 
equation
\[
\Gamma=\gamma\exp \Bigl (-\frac{u}{k_{\rm B}T}\Bigr ),
\]
where $\gamma$ stands for the attempt rate,
$T$ is the absolute temperature,
and $k_{\rm B}$ denotes the Boltzmann constant.
For isothermal processes at a fixed temperature $T$,
we introduce the dimensionless energy $v=u/(k_{\rm B}T)$ and
find that
\begin{equation}\label{3-5}
\Gamma=\gamma\exp (-v).
\end{equation}
Distribution of meso-domains with various activation energies $v$
is described by the quasi-Gaussian formula 
\begin{equation}\label{3-6}
f(v)=f_{0}\exp \Bigl (-\frac{v^{2}}{2\Sigma^{2}}\Bigr )
\quad
(v\geq 0),
\end{equation}
where $f$ stands for the distribution function.
This function is characterized by the only parameter $\Sigma>0$ 
that serves as a measure of inhomogeneity of a polymer network \cite{DG03a}.
The pre-factor $f_{0}$ is determined from the normalization condition
\begin{equation}\label{3-7}
\int_{0}^{\infty} f(v) {\rm d}v=1.
\end{equation}
According to Equation (\ref{3-5}), the average rate of rearrangement of 
temporary bonds in a transient network reads
\begin{equation}\label{3-8}
\bar{\Gamma}=\gamma \int_{0}^{\infty} \exp(-v) f(v) {\rm d}v.
\end{equation} 

Focusing on experimental data in uniaxial tests,
we disregard volume deformation and treat of HDPE as 
an incompressible medium
(according to \cite{Dro10,DCK10}, the Poisson ratio of 
polyethylene $\nu$ belongs to the interval between 0.48 and 0.49).

Constitutive equations for a semicrystaline polymer under an arbitrary
three-dimensional deformation with small strains were derived in \cite{DHC23}
by means of the Clausius--Duhem inequality.
Under uniaxial tensile relaxation with a fixed strain $\epsilon_{0}$, 
the governing relation reads
\begin{equation}\label{3-31}
\sigma(t) = \sigma_{0}
\Bigl [ 1 -\kappa \int_{0}^{\infty} f(v)
\Bigl (1-\exp(-\Gamma(v) t)\Bigr ){\rm d}v \Bigr ],
\end{equation}
where $\sigma(t)$ denotes tensile stress at an arbitrary instant $t\geq 0$,
\begin{equation}\label{3-33}
\sigma_{0}=E \epsilon_{0}
\end{equation}
stands for the stress at the initial instant $t=0$,
and $\kappa$ denotes the ratio of the number of active chains
to the total number of (active and permanent) chains per unit volume.

Given a strain $\epsilon_{0}$, Equations (\ref{3-31}), (\ref{3-33}) together 
with Equation (\ref{3-5}) for $\Gamma(v)$ and Equation (\ref{3-6}) for $f(v)$ 
involve four material parameters:
(i) the Young's modulus $E$,
(ii) the rate of separation of active chains from their junctions
$\gamma$,
(iii) the measure of inhomogeneity of the equivalent network $\Sigma$,
and (iv) the fraction of temporary bonds in the network $\kappa$.
These quantities are found in Section 4 by matching experimental 
data on virgin and recycled HDPE.

Under uniaxial tensile creep with a fixed stress $\sigma$, 
the tensile strain $\epsilon$ is given by
\begin{equation}\label{3-37}
\epsilon=\epsilon_{\rm e}+\epsilon_{\rm pa}+\epsilon_{\rm pc},
\end{equation}
where $\epsilon_{\rm e}$ stands for the strain caused by 
the viscoelastic response (rearrangement of temporary bonds 
between polymer chains in the network),
and $\epsilon_{\rm pa}$, $\epsilon_{\rm pc}$ are the strains induced
by the viscoplastic flows (sliding of junctions between chains) 
in the amorphous and crystalline regions, respectively.

The strain $\epsilon_{\rm e}(t)$ is determined by the formula
\begin{equation}\label{3-34}
\epsilon_{\rm e}(t)=\frac{\sigma}{E}+\kappa \int_{0}^{\infty} f(v) z(t,v) {\rm d}v,
\end{equation}
where the function $z(t,v)$ obeys the differential equation
\begin{equation}\label{3-36}
\frac{\partial z}{\partial t}=\Gamma(v)( \epsilon_{\rm e}-z),
\qquad
z(0,v)=0.
\end{equation}

The viscoplastic flow in the amorphous phase is governed by 
the equation
\begin{equation}\label{3-40}
\epsilon_{\rm pa}(t)=A \Bigl [1-\exp(-\alpha \sqrt{t})\Bigr ],
\end{equation}
where $A$ and $\alpha$ are adjustable parameters.
The coefficient $A$ stands for the maximum viscoplastic strain 
induced by sliding of junctions between chains, 
and $\alpha$ characterizes the rate of sliding of junctions 
with respect to their initial positions.
An analog of Equation (\ref{3-40}) (with $\sqrt{t}$ replaced with $t$)
is conventionally used to describe at the initial stage of flow 
of dislocations in crystalline materials \cite{WBW10}.

The viscoplastic flow in the crystalline phase is described by 
the Norton equation \cite{SOK22}
\begin{equation}\label{3-38}
\epsilon_{\rm pc}(t)=B t,
\end{equation}
where $B$ is an adjustable parameter.

Given a stress $\sigma$, the viscoplastic flow under creep conditions 
is determined by three material parameters:
(i) the rate of sliding of junctions in the amorphous phase with
respect to their initial positions $\alpha$,
(ii) the maximum viscoplastic strain $A$ induced by sliding 
of junctions in the amorphous phase,
and 
(iii) the rate of plastic flow in the crystalline phase $B$.

The influence of stress on the rate of sliding of junctions
in amorphous regions $\alpha$ is described the power-law relation
\begin{equation}\label{3-42}
\alpha=\alpha_{0}
\qquad
(\sigma<\sigma_{\ast}),
\qquad
\alpha=\alpha_{0}+\alpha_{1}(\sigma-\sigma_{\ast})^{m},
\end{equation}
where $\alpha_{0}$, $\alpha_{1}$ and $m$ are material parameters,
and $\sigma_{\ast}$ denotes the threshold stress below 
which the viscoplastic flow is not observed.

The effect of stress $\sigma$ on the coefficient $A$ is determined
by the formula
\begin{equation}\label{3-41}
A=\frac{A_{1}}{1+\exp[a(\sigma-\sigma_{\ast})]}+A_{2},
\end{equation}
where $A_{1}$, $A_{2}$ and $a$ are material parameters.
Equation (\ref{3-41}) implies that $A$ adopts different values 
$A_{1}+A_{2}$ and $A_{2}$ far below and far above some threshold 
stress $\sigma_{\ast}$.
Transition from the limiting values occurs in the close vicinity 
of $\sigma_{\ast}$ only.
The rate of transition is characterized by the coefficient $a$.

The influence of stress $\sigma$ on the rate of viscoplastic flow in
the crystalline phase $B$ is governed by the equation
\begin{equation}\label{3-39}
B=0
\qquad
(\sigma<\sigma_{\ast}),
\qquad
B= B_{1} (\sigma-\sigma_{\ast})^{n}
\qquad
(\sigma\geq \sigma_{\ast}),
\end{equation}
where $B_{1}$ and $n$ are adjustable parameters.
The coefficient $\sigma_{\ast}$ denotes the threshold stress below 
which the viscoplastic flow in spherulites does not occur.

Equations (\ref{3-42}) and (\ref{3-39}) imply that sliding of 
junctions is weakly affected by tensile stresses when 
these stresses remain relatively small, but the rate of sliding
increases pronouncedly when $\sigma$ exceeds its threshold 
value $\sigma_{\ast}$.

\section{Results and Discussion}

\subsection{Fitting of experimental data}

Material parameters for recycled HDPE are determined by matching
observations presented in Figures 2 to 4.

\subsubsection{Relaxation tests}

We begin with the analysis of experimental data in relaxation 
tests.
Each set of data in Figure 2 is fitted separately by means of 
Equation (\ref{3-31}),
where $\Gamma(v)$ is given by  Equation (\ref{3-5}),
$f(v)$ is determined by Equation (\ref{3-6}), 
and $\sigma_{0}$ is found from Equation (\ref{3-33}).
Given $\gamma$ and $\Sigma$, the coefficients $\sigma_{0}$ and $\kappa$
in Equation (\ref{3-31}) are calculated by using the least-squares technique.
The parameters $\gamma$ and $\Sigma$ are determined by the nonlinear
regression method from the condition of minimum for the expression
\[
\sum_{k} \Bigl [\sigma_{\rm exp}(t_{k})-\sigma_{\rm sim}(t_{k})\Bigr ]^{2},
\]
where summation is performed over all instants $t_{k}$ at which the data
are reported,
$\sigma_{\rm exp}$ denotes the tensile stress measured in the
corresponding relaxation test,
and $\sigma_{\rm sim}$ is given by Equstion (\ref{3-31}).
\vspace*{3 mm}

\noindent
{\bf Table 2:} Material parameters for recycled HDPE 
(relaxation tests with various strains $\epsilon_{0}$)
\vspace*{3 mm}

\begin{center}
\begin{tabular}{@{} c r c c c @{}}\hline
$\epsilon_{0}$ &  $\sigma_{0}$ MPa & $\gamma$ s  & $\Sigma$  & $\kappa$   
\cr \hline
0.01       & 12.99             & 1.5         & 5.6       &  0.810   \\
0.02       & 16.91             & 1.1         & 5.6       &  0.770   \\
0.05       & 22.37             & 0.9         & 5.7       &  0.711   \\
\hline
\end{tabular}
\end{center}
\vspace*{3 mm}

Figure 2 demonstrates good agreement between the experimental data
in relaxation tests and their approximation of the model with 
the material constants collected in Table 2.
This table shows that the coefficients $\gamma$, $\Sigma$ and $\kappa$
are weakly affected by tensile strain $\epsilon_{0}$
($\gamma$ and $\kappa$ are reduced slightly with $\epsilon_{0}$,
whereas $\Sigma$ remains practically constant).
In what follows, changes in these coefficients with $\epsilon_{0}$
are disregarded.
Taking the value $\sigma_{0}$ at $\epsilon_{0}=0.01$ from Table 2 
and using Equation (\ref{3-33}), we calculate the Young's modulus  
$E=1.299$ GPa. 
This value is very close to the value $\bar{E}=1.305$ GPa found
by approximation of the stress--strain curves under tensile deformation
(Figure 1) and reported in Table 1.

Fitting observations in relaxation tests on virgin HDPE was performed
in \cite{DHC23}.
For comparison, the best-fit values of $E$, $\gamma$, $\Sigma$, and $\kappa$ 
for virgin and recycled HDPE are collected in Table 3.
\vspace*{3 mm}

\noindent
{\bf Table 3:} Material parameters for virgin and recycled HDPE 
(relaxation tests with strain $\epsilon_{0}=0.01$)
\vspace*{3 mm}

\begin{center}
\begin{tabular}{@{} l c c c c @{}}\hline
HDPE &  $E$ GPa & $\gamma$ s  & $\Sigma$  & $\kappa$   
\cr \hline
Virgin     & 0.947              & 2.1         & 7.2       &  0.862   \\
Recycled   & 1.299              & 1.5         & 5.6       &  0.810   \\
\hline
\end{tabular}
\end{center}

Table 3 shows that all parameters characterizing the linear viscoelastic
behavior of virgin and recycled HDPE adopt similar values.
The Young's modulus $E$ of the recycled HDPE is higher (by 37\%),
whereas its attempt rate $\gamma$,
measure of inhomogeneity of the temporary network $\Sigma$, 
and fraction of temporary junctions in the network $\kappa$ 
are lower (by 29, 22, and 6 \%, respectively).
This difference may be explained by a higher degree of crystallinity
of the recycled HDPE \cite{GJH03,HKO20}, as well as a higher degree 
of long-chain branching in this polymer caused by its degradation 
under in-service conditions and recycling \cite{PCC04}.
The average rates of rearrangement of temporary bonds $\bar{\Gamma}$
(determined by Equation (\ref{3-8}))
coincide practically for virgin and recycled HDPE and 
read $\bar{\Gamma}=0.255$ and $\bar{\Gamma}=0.231$ s$^{-1}$, respectively.

\subsubsection{Creep tests}

We now approximate the experimental creep diagrams on recycled HDPE depicted
in Figures 3 to 5.
Each set of data is matched separately by using the following algorithm.

\input{figure-06.tex}
\input{figure-07.tex}
\input{figure-08.tex}

Given a stress $\sigma$, the viscoelastic strain $\epsilon_{\rm e}$ 
is determined from Equations (\ref{3-34}) and (\ref{3-36}).
These equations are integrated numerically by the Runge--Kutta
method with the material parameters reported in Table 3.
Afterwards, the viscoplastic strain 
$\epsilon_{\rm p}=\epsilon_{\rm pa}+\epsilon_{\rm pc}$
is determined from Equations (\ref{3-40}) and (\ref{3-38}).
Given $\alpha$, the coefficients $A$ and $B$ are found by the least-squares
technique.
The parameter $\alpha$ is calculated by the nonlinear regression method
from the condition of minimum for the expression
\[
\sum_{k} \Bigl [\epsilon_{\rm p\;exp}(t_{k})-\epsilon_{\rm p\;sim}(t_{k})\Bigr ]^{2},
\]
where summation is performed over all instants $t_{k}$ at which the data
are reported,
$\epsilon_{\rm p\; exp}$ is determined from Equation (\ref{3-37}),
and $\epsilon_{\rm p\; sim}$ is found from Equations
(\ref{3-40}) and (\ref{3-38}).

Figures 3 and 4 demonstrate good agreement between results of numerical 
analysis and the experimental creep diagrams along the intervals of 
primary and secondary creep.
Slight deviations between the data and results of simulation
are observed along the intervals of tertiary creep in Figures 4 and 5
only.
These discrepancies arise because the model does not account for damage 
accumulation at the final stage of creep flow.

The coefficients $\alpha$, $A$, and $B$ for recycled HDPE are plotted 
versus tensile stress $\sigma$ in Figures 6 to 8.
For comparison, we present also the corresponding dependencies for
virgin HDPE obtained in \cite{DHC23}.
The data are approximated by Equations (\ref{3-42})--(\ref{3-39})
with the adjustable parameters reported in Table 4.
The coefficients $\alpha_{0}$, $\alpha_{1}$, $A_{1}$ and $B_{1}$ are 
calculated by the least-squares technique.
The parameters $m$, $n$ and $a$ are determined by the nonlinear regression
method.
\vspace*{3 mm}

\noindent
{\bf Table 4:} Material parameters for virgin and recycled HDPE 
(creep tests with various stresses $\sigma$)

\begin{center}
\begin{tabular}{@{} l c c c c @{}}\hline
Coefficient $\alpha$\\
\hline
HDPE &  $\alpha_{0}$ & $\alpha_{1}$ & $\sigma_{\ast}$ MPa & $m$   
\cr \hline
Virgin     & 1.982              & $8.08\cdot 10^{-7}$     & 9.0       &  7.0   \\
Recycled   & 0.0                & $2.34\cdot 10^{-3}$     & 9.2       &  4.0   \\
\hline
Coefficient $A$\\
\hline
HDPE &  $A_{0}$ & $A_{1}$ & $\sigma_{\ast}$ MPa & $a$ MPa$^{-1}$
\cr \hline
Virgin     & $4.67\cdot 10^{-2}$    & $-3.90\cdot 10^{-2}$    & 13.5       &  1.1   \\
Recycled   & $2.90\cdot 10^{-2}$    & $2.01\cdot 10^{-2}$     & 16.0       &  4.0   \\
\hline
Coefficient $B$\\
\hline
HDPE &  $B_{1}$ & $\sigma_{\ast}$ MPa & $n$ 
\cr \hline
Virgin     & $3.51\cdot 10^{-9}$   & 9.0       &  8.0   \\
Recycled   & $1.28\cdot 10^{-7}$   & 9.0       &  6.6   \\
\hline
\end{tabular}
\end{center}
\vspace*{3 mm}

Figures 6 and 8 demonstrate similar effects of tensile stress $\sigma$
on the rates $\alpha$ and $B$ of the viscoplastic flow in amorphous and 
crystalline regions.
It should be noted, however, that the growth of $\alpha$ and $B$ with $\sigma$
is stronger (the quantities $m$ and $n$ are higher) in virgin HDPE compared 
with recycled HDPE.

Figure 7 shows a difference between the effects of stress $\sigma$
on the coefficient $A$ for virgin and recycled HDPE.
The parameter $A$ increases with $\sigma$ in the virgin polymer
and decreases with tensile stress in the recycled HDPE.
However, the stress-induced changes in $A$ are not substantial.
This coefficient varies in the interval between 0.01 and 0.05 for
virgin HDPE and in the interval between 0.03 and 0.05 for its recycled 
analog.

\subsection{Validation of the model}

To predict time to failure $t_{\rm f}$ of recycled HDPE in medium-term
creep tests (with the duration of several days),
we integrate Equations (\ref{3-34}) and (\ref{3-36}) with
the parameters $E$, $\gamma$, $\Sigma$ and $\kappa$
reported in Table 3 and the parameters $\alpha$, $A$ and $B$
determined by Equations (\ref{3-42})--(\ref{3-39}) with 
the coefficients listed in Table 4.
For each stress $\sigma$, evolution of tensile strain $\epsilon$ with
time $t$ is determined by Equation (\ref{3-37}).
Time to failure $t_{\rm f}$ is calculated from the condition
\begin{equation}\label{4-1}
\epsilon(t_{\rm f})=\epsilon_{\ast},
\end{equation} 
where the critical strain $\epsilon_{\ast}=0.4$ is found from 
Figures 4 and 5 (it corresponds to transition to the tertiary
creep in short-term creep tests).
Equation (\ref{4-1}) is based on the conventional assumption that 
the duration of the interval of tertiary creep is negligible compared 
with the durations of intervals of primary and secondary creep.


\thicklines
\setlength{\unitlength}{0.5 mm}
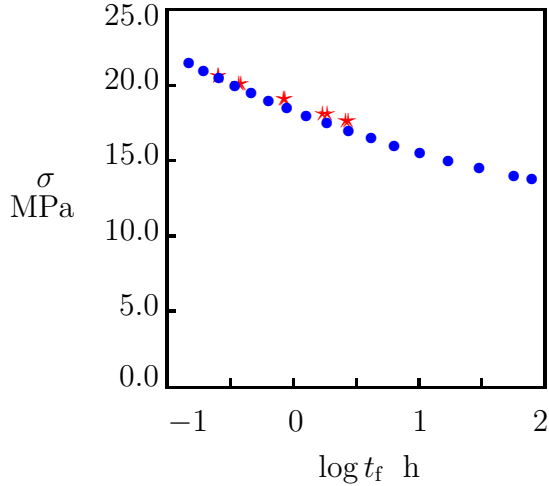
\begin{figure}[!tbh]
\begin{center}
\begin{picture}(100,100)
\put(0,0){\framebox(100,100)}
\multiput(16.67,0)(16.67,0){5}{\line(0,1){2}}
\multiput(0,20)(0,20){4}{\line(1,0){2}}
\put(0,-12){$-1$}
\put(32,-12){0}
\put(65,-12){1}
\put(97,-12){2}
\put(40,-24){$\log t_{\rm f}$\hspace*{1 mm} h}

\put(-13,0){0.0}
\put(-13,19){5.0}
\put(-17,39){10.0}
\put(-17,59){15.0}
\put(-17,79){20.0}
\put(-17,96){25.0}
\put(-35,53){$\sigma$}
\put(-42,45){MPa}

\color{red}
\put(  45.84,   68.70){$\star$} 
\put(  45.19,   68.70){$\star$} 
\put(  39.00,   70.70){$\star$} 
\put(  40.30,   70.70){$\star$} 
\put(  28.70,   74.70){$\star$} 
\put(  29.03,   74.70){$\star$} 
\put(  16.84,   78.70){$\star$} 
\put(  17.32,   78.70){$\star$} 
\put(  11.29,   80.70){$\star$} 
\put(  11.29,   80.70){$\star$} 
\color{blue}
\put(   5.59,   86.00){\circle*{2.5}}
\put(   9.46,   84.00){\circle*{2.5}}
\put(  13.49,   82.00){\circle*{2.5}}
\put(  17.71,   80.00){\circle*{2.5}}
\put(  22.12,   78.00){\circle*{2.5}}
\put(  26.75,   76.00){\circle*{2.5}}
\put(  31.62,   74.00){\circle*{2.5}}
\put(  36.75,   72.00){\circle*{2.5}}
\put(  42.18,   70.00){\circle*{2.5}}
\put(  47.93,   68.00){\circle*{2.5}}
\put(  53.98,   66.00){\circle*{2.5}}
\put(  60.20,   64.00){\circle*{2.5}}
\put(  66.91,   62.00){\circle*{2.5}}
\put(  74.46,   60.00){\circle*{2.5}}
\put(  82.76,   58.00){\circle*{2.5}}
\put(  91.88,   56.00){\circle*{2.5}}
\put(  96.72,   55.02){\circle*{2.5}}

\end{picture}
\end{center}
\vspace*{10 mm}

\caption{Tensile stress $\sigma$ versus time-to-failure $t_{\rm f}$
for recycled HDPE under creep conditions.
Asterisks: experimental data in creep tests with various stresses
$\sigma$.
Circles: predictions of the model.}
\end{figure}

Results of numerical analysis are reported in Figure 9 where 
the stress $\sigma$ is plotted versus time to failure $t_{\rm f}$.
Results of simulation (circles) and presented together with the experimental 
data (asterisks) in creep tests with stresses $\sigma=17.5$, 18.0, 19.0, 
20.0 and 20.5 MPa (Figure 4).
Figure 9 confirms the ability of the model to predict the stress-time 
to failure diagrams on recycled HDPE in medium-term creep tests.

\subsection{Lifetimes of virgin and recycled HDPE}

To compare the lifetimes of virgin and recycled HDPE under creep
conditions, their times to failure $t_{\rm f}$ in long-term
creep tests (with the duration up to 100 years)
are calculated numerically.
For this purpose, Equations (\ref{3-34}) and (\ref{3-36}) 
are integrated with the parameters $E$, $\gamma$, $\Sigma$ and $\kappa$
listed in Table 3 and the parameters $\alpha$, $A$ and $B$
determined by Equations (\ref{3-42})--(\ref{3-39}) with 
the coefficients collected in Table 4.
For each polymer under consideration, time to failure $t_{\rm f}$ 
is found from condition (\ref{4-1}).
\vspace*{5 mm}

\input{figure-10.tex}

Results of simulation are depicted in Figure 10 where tensile stress
$\sigma$ is plotted versus time-to failure $t_{\rm f}$
(d, m, and y stand for day, month and year, respectively).
In this figure, circles denote results of numerical analysis, 
and solid lines provide their approximations by the power-law equation 
\cite{Gue06,BLW22}
\begin{equation}\label{4-2}
\sigma=\sigma_{0}+\sigma_{1}\Bigl (\frac{t_{\rm f0}}{t_{\rm f}}\Bigr )^{\delta}
\end{equation}
with the characteristic time $t_{\rm f0}=1$ h.
The best-fit values of the material constants
$\sigma_{0}$, $\sigma_{1}$ and $\delta$ in Equation (\ref{4-2})
are collected in Table 5.

Figure 10 provides the main result of this study.
It shows that virgin and recycled HDPE with similar mechanical 
characteristics (the Young modulus, yield stress,
and strain at break) have practically identical lifetimes under
tensile creep conditions.
Unlike previous observations that revealed a pronounced decay in creep
resistance when the same polymer was recycled \cite{APB15,SSM22},
our analysis shows that that creep resistance of the recycled material
(prepared from post-consumer plastic waste of a different grade of HDPE)
is similar to that of the virgin HDPE.
Figure 10 demonstrates that specimens withstand 50 years of creep under 
tensile stresses lower than 10.9 and 10.3 MPa for virgin and recycled HDPE, 
respectively.
Replacement of the virgin HDPE with its recycled analog
leads to an insignificant reduction in the critical stress by 5.5\% only.
This may be explained by a higher stiffness of the recycled polymer
(Table 3) and a less pronounced growth of the rates of viscoplastic flow
in amorphous (Figure 6) and crystalline (Figure 8) regions.

\noindent
{\bf Table 5:} Material parameters for virgin and recycled HDPE 
(long-term creep tests with various stresses $\sigma$)
\vspace*{3 mm}

\begin{center}
\begin{tabular}{@{} l c c c @{}}\hline
HDPE &  $\sigma_{0}$ MPa & $\sigma_{1}$ MPa  & $\delta$ 
\cr \hline
Virgin     & 9.609              & 9.108      & 0.129   \\
Recycled   & 9.241              & 9.075      & 0.155   \\
\hline
\end{tabular}
\end{center}
\vspace*{3 mm}

Although the above conclusion opens a new area of engineering applications
for recycled polymers, it should be treated with caution.
It is conventionally presumed that the stress--time to failure diagrams 
for polyethylene pipes consist of three intervals \cite{ZLH22}.
The model under consideration describes adequately the lifetime 
along the first interval (which corresponds to ductile failure of HDPE).
The other intervals are associated with quasi-brittle failure driven
by slow growth of micro-cracks and brittle failure caused by chemical 
degradation of this polymer \cite{SOK22}.
Analysis of crack resistance of virgin and recycled HDPE under 
various loading programs will be conducted in a subsequent study.

\section{Conclusions}

Replacement of virgin polymers with their recycled analogs
is attractive from the economical and environmental standpoints,
but it is not always possible because of substantial deterioration 
of the mechanical properties of recyclates induced by their 
degradation during the service life and recycling process.
Due to a pronounced reduction in resistance to creep, fatigue, and 
crack growth, recycled polymers cannot met the same property 
profiles for long-term performance applications as their virgin 
counterparts.

To avoid a noticeable decline in mechanical characteristics
of recycled materials, it was proposed in \cite{IJM21,JDR21} 
to replace a virgin polymer of a special grade with its recycled 
analog (a post-consumer recyclate prepared from another grade) 
whose properties after recycling (evaluated in tensile tests) 
are close to those of the virgin material.
This approach opens a way for the use of recycled polymers 
in short-term application.
However, its suitability for long-term applications has not 
yet been evaluated.

A thorough experimental investigation is performed of the mechanical 
response of virgin HDPE used for insulation of pipes 
for district heating and recycled HDPE manufactured from post-consumer 
household plastic waste.
The Young's modulus, yield stress, and elongation to break of these
two polymers adopt similar values (Table 1).
Their viscoelastic and viscoplastic responses in short-term tensile 
relaxation and creep tests are adequately described by the same model 
presented in Section 3 (Figures 2 to 5).
However, the material parameters of virgin and recycled HDPE
determined by fitting experimental data in relaxation and creep tests
differ noticeably (Tables 3 and 4).

The lifetime of virgin and recycled HDPE under creep conditions is
evaluated by means of numerical simulation of the governing equations
with the material parameters found by matching observations in
short-term tests.
Validation of the model is performed by comparison of its predictions
with experimental data in short-term creep tests (Figure 9).
The main result of the study is that the stress--time to failure
diagrams for virgin and recycled HDPE practically coincide
(Figure 10 and Table 5).
Numerical analysis reveals that recycled HDPE withstands 50 years 
of creep under tensile stress which is lower by only 5.5\% than that
for virgin HDPE.

\subsection*{Acknowledgement}

Financial support by Innovationsfonden (Innovation Fund Denmark, 
project 9091-00010B) is gratefully acknowledged.

\end{document}